\documentclass[12pt]{article}

\usepackage{graphics}
\usepackage{color}
\usepackage{amssymb,amsmath}
\usepackage{slashed}
\usepackage{epstopdf}
\usepackage{booktabs}
\usepackage{mathrsfs}
\usepackage[export]{adjustbox}
\usepackage{epsfig}
\usepackage{graphicx,color}

\newcommand{\bea}{\begin{eqnarray}}
\newcommand{\eea}{\end{eqnarray}}

\numberwithin{equation}{section}

\begin{document}
\begin{titlepage}
%
%
\vspace*{10mm}
\begin{center}
\baselineskip 25pt 
{\Large\bf
Running Non-Minimal Inflation with Stabilized Inflaton Potential
}
\end{center}
\vspace{5mm}
\begin{center}
{\large
Nobuchika Okada\footnote{okadan@ua.edu}
and
Digesh Raut\footnote{draut@crimson.ua.edu}
}
\end{center}
\vspace{2mm}

\begin{center}
{\it
Department of Physics and Astronomy, University of Alabama, \\
Tuscaloosa, Alabama 35487, USA
}
\end{center}
\vspace{0.5cm}
\begin{abstract}
In the context of the Higgs model involving gauge and Yukawa interactions 
   with the spontaneous gauge symmetry breaking,  
   we consider $\lambda \phi^4$ inflation with non-minimal gravitational coupling, 
   where the Higgs field is identified as inflaton.   
Since the inflaton quartic coupling is very small, 
   once quantum corrections through the gauge and Yukawa interactions are taken into account, 
   the inflaton effective potential most likely becomes unstable.  
In order to avoid this problem, we need to impose stability conditions 
   on the effective inflaton potential, which lead to not only non-trivial relations 
   amongst the particle mass spectrum of the model, but also correlations  
   between the inflationary predictions and the mass spectrum. 
For concrete discussion, we investigate the minimal $B-L$ extension of the Standard Model 
   with identification of the $B-L$ Higgs field as inflaton. 
The stability conditions for the inflaton effective potential fix the mass ratio 
   amongst the $B-L$ gauge boson, the right-handed neutrinos and the inflaton. 
This mass ratio also correlates with the inflationary predictions. 
In other words, if the $B-L$ gauge boson and the right-handed neutrinos are discovered in future,  
   their observed mass ratio provides constraints on the inflationary predictions. 

\end{abstract}
\end{titlepage}

\section{Introduction}
Current understanding about the origin of our universe is that, for a very brief moment at the beginning, 
   our universe went through a period of rapid accelerated expansion known as inflation. 
Inflation scenario \cite{inflation1, inflation2, chaotic_inflation, inflation3}
   was originally proposed to solve serious problems in the Standard Big-Bang Cosmology, 
   namely, the horizon, flatness and monopole problems. 
In addition and more importantly in the view point of the current cosmological observations, 
   inflation provides a mechanism to create primordial density fluctuations 
   of the early universe which seed the formation of large scale structure of the universe that we see today.  
In a simple inflation scenario, inflation is driven by a single scalar field (inflaton) that slowly rolls down to 
   its potential minimum (slow-roll inflation).  
During the slow-roll era, the inflaton energy is dominated by a slowly varying potential, 
   which causes the universe to undergo a phase of an accelerated expansion. 
Quantum fluctuations of the inflaton field are stretched to macroscopic scales by inflation 
   to yield the primordial density  fluctuations. 
After inflation, the inflaton decays to the Standard Model (SM) particles and 
   the decay products heat up the universe (reheating). 
The success of big bang nucleosynthesis scenario 
   requires the reheating temperature to be $T_{R} \gtrsim 1$ MeV.

Recently the Planck 2015 results \cite{Planck2015} have set an upper bound on the tensor-to-scalar ratio as $r \lesssim 0.11$ 
    while the best fit value for the spectral index ($n_s$) is $0.9655 \pm 0.0062$ at $68 \%$ CL. 
Hence, the simple chaotic inflationary scenario with the inflaton potentials $V \propto  \phi^4$ and $V \propto \phi^2$ 
    are disfavored because their predictions for $r$ are too large. 
Among many inflation models, $\lambda \phi^4$ inflation with non-minimal gravitational coupling 
    ($\xi \phi^2 {\cal R}$, where $\phi$ is inflaton, ${\cal R}$ is the scalar curvature, and $\xi$ is a dimensionless coupling) 
    is a very simple model, which can satisfy the constraints by the Planck 2015 
    with $\xi \gtrsim 0.001$ \cite{NonMinimalUpdate}.

Given that we need interactions between SM particles and inflaton for a successful reheating of the universe, 
    a more compelling inflation scenario would be where the inflaton field plays another important role 
    in particle physics.    
As an example of such a scenario, we may consider the (general) Higgs model, where a scalar (Higgs) field  plays
    the crucial role to spontaneously break the gauge symmetry of the model, and we identify the Higgs field as inflation. 
The SM Higgs inflation \cite{Higgs_inflation1, Higgs_inflation2, Higgs_inflation3} is nothing but this scenario, 
    where the SM Higgs boson plays the role of inflaton with non-minimal gravitational coupling.  
Because of the observed Higgs boson mass of around 125 GeV, 
    the SM Higgs effective potential is likely unstable.\footnote{
The stability of SM Higgs potential is very sensitive to the initial value of top quark pole mass, 
   and we need more precise measurements for it \cite{instability1}. 
See also \cite{instability2} for the effect of possible Planck scale physics to the effective Higgs potential. 
}
If this is the case, the original SM Higgs inflation cannot work any more, and
  some extension is necessary \cite{OS}.\footnote{
Supersymmetric version of the Higgs inflation \cite{SHI1, SHI2, SHI3, SHI4} 
   is free from this instability problem because of supersymmetry.   
It has been shown in \cite{KimKawai} that a large non-Gaussianity can be generated in this class of models. 
}     
However, we may apply the same idea to the general Higgs model and identify the Higgs field  
   of the model (not the SM Higgs field) as inflaton in the presence of non-minimal gravitational coupling.  
For a simple example, see \cite{ORS-BL}.

As in the SM, the Higgs field in the general Higgs model has the gauge, Yukawa and quartic Higgs interactions. 
For complete analysis of inflation scenario in the Higgs model, we consider the effective inflaton/Higgs potential 
   by taking quantum corrections into account. 
In fact, we see that quantum corrections most likely cause an instability of the effective inflaton potential. 
Note that unless the non-minimal coupling ($\xi$) is very large, the quartic inflaton coupling 
   is very small \cite{NonMinimalUpdate}.  
Hence, quantum corrections to the effective potential are dominated by the gauge and Yukawa interactions. 
We consider the renormalization group (RG) improved effective potential described as 
\bea
   V(\phi) = \frac{1}{4} \lambda(\phi) \;  \phi^4, 
\eea
    where $\phi$ denotes inflaton, and  $\lambda(\phi)$ is the running quartic coupling
    satisfying the (one-loop) RG equation of the form,  
\bea 
  16 \pi^2 \frac{d \lambda}{d \ln \mu} \simeq  C_g \; g^4  -  C_Y \; Y^4 . 
\eea  
Here, $g$ and $Y$ are the gauge and Yukawa couplings, respectively, 
    and $C_g$ and $C_Y$ are positive coefficients 
    whose actual values are calculable once the particle contents of the model are defined.   
Since the quartic coupling is very small, we have neglected terms proportional to $\lambda$ 
    ($\lambda^2$ term and the anomalous dimension term).  
The solution to the RG equation is controlled by $g$ and $Y$, 
   which are much larger than $\lambda$ and independent of $\lambda$.  
Therefore, we expect that unless the beta function is extremely small, the running inflaton quartic coupling $\lambda$ 
   is driven to be negative in the vicinity of the inflation initial value,  
   in other words, the effective potential has true minimum (far) away from the vacuum set 
   by the Higgs potential at the tree-level.

A simple way to avoid this problem is to require the beta function to vanish at the initial inflaton value
   (the stationary condition of $\lambda$ with respect to $\phi$), namely, $C_g \;  g - C_Y \;  Y = 0 $.  
This condition leads to a relation between $g$ and $Y$, equivalently, 
   a mass relation between the gauge boson and fermion in the Higgs model. 
Since the Higgs quartic coupling at low energy is evaluated by solving the RG equation, 
   in which the gauge and Yukawa couplings dominate,  
   the resultant Higgs mass also has a relation to the gauge and fermion masses.  
The stability of the effective potential also requires the positivity of the second derivative of the potential, 
   which leads to another constraint on the gauge and Yukawa couplings.  
In the slow-roll inflation, the inflationary predictions are determined by the slow-roll parameters 
   defined with the potential and its derivatives, and therefore, 
   the inflationary predictions have a correlation with the mass spectrum of the Higgs model.

In order to explicitly show the mass relation and the correlation between the particle mass spectrum and 
   inflationary predictions, we take the minimal $B-L$ model as an example. 
This model is a very simple, well-motivated extension of the SM, 
   where the global $B-L$ (baryon number minus lepton number) in the SM is gauged.  
Three right-handed neutrinos and the $B-L$ Higgs field (which is identified as inflaton) 
   are introduced for the cancellation of the gauge and gravitational anomaly 
   and the $B-L$ gauge symmetry breaking, respectively.  
Associated with the $B-L$ gauge symmetry breaking, the $B-L$ gauge boson 
   and the right-handed neutrinos acquire their masses. 
With the generation of the Majorana right-handed neutrino masses, 
   the seesaw mechanism \cite{Seesaw} for the light neutrino mass generation is automatically 
   implemented in this model.  
Analyzing the RG evolutions of the $B-L$ sector and the effective inflaton ($B-L$ Higgs) potential, 
   we show the particle mass spectrum and its correlation to the inflationary predictions.  
Through the correlation, the Planck 2015 results provide us with constraints on the particle mass spectrum.

This paper is organized as follows. 
In the next section, we briefly review the $\lambda \phi^4$ inflation with non-minimal gravitational coupling at the tree-level,  
   and discuss the inflationary predictions  in the light of the Planck 2015 results.     
In Sec.~3, we introduce the minimal $B-L$ extension of the SM and calculate the RG improved effective Higgs potential. 
We show the particle mass spectrum derived from the stability conditions of the effective potential 
   and its correlation to the inflationary predictions. 
We then compare our results for various values of the non-minimal gravitational coupling $\xi$ 
   with the Planck 2015 results.     
In Sec.~5, we discuss reheating scenario in the $B-L$ Higgs inflation for the completion of our inflationary scenario. 
Sec.~6 is devoted to conclusions.

\section{Non-minimal $\lambda \phi^4$ inflation at tree-level}
In the Jordan frame, the action of our inflation model is given by 
   (hereafter we always work in the Planck unit, $M_P= M_{Pl}/\sqrt{8 \pi}=1$, 
    where $M_{Pl} =1.22 \times 10^{19}$ GeV is the Planck mass)
\bea
\mathcal{S}_J= \int \mathrm{d}^4 x  \sqrt{-g} \left[-\frac{1}2 f(\phi) \mathcal{R}  
   +\frac{1}2(\nabla \phi)^2 - V(\phi) \right]  ,
\eea
   where $f(\phi) = (1 +\xi \phi^2)$ with $\xi$ being a positive, dimensionless parameter, 
   and the inflaton potential is 
\bea
   V(\phi)= \frac{1}{4} \lambda \; \phi^4. 
\eea     
Using the conformal transformation,  $ g_{E \mu\nu} = f(\phi) g_{\mu\nu}$,  
   the action in the Einstein frame is descried as  
\bea
\mathcal{S}_{E}= \int \mathrm{d}^4 x  
 \sqrt{-g_E} \left[-\frac{1}2 \mathcal{R}_E  +\frac{1}2 \left( \nabla_E \sigma \right)^2 
      - \frac{V[\phi(\sigma)]}{f^2[\phi(\sigma)]}  \right] .
\eea
In the Einstein frame with a canonical gravity sector, 
  we describe the theory with a new inflaton field ($\sigma$) which has a canonical kinetic term. 
The relation between $\sigma$ and the original inflaton field $\phi$ is given by  
\bea
  (\sigma^\prime(\phi))^2 = \frac{1+(1+6\xi)\xi \phi^2}{(1 +\xi \phi^2)^2} \ ,
\eea
  where a prime denotes the derivative with respect to $\phi$. 
  
\begin{figure}[h]
\begin{center}
\includegraphics[scale=0.85]{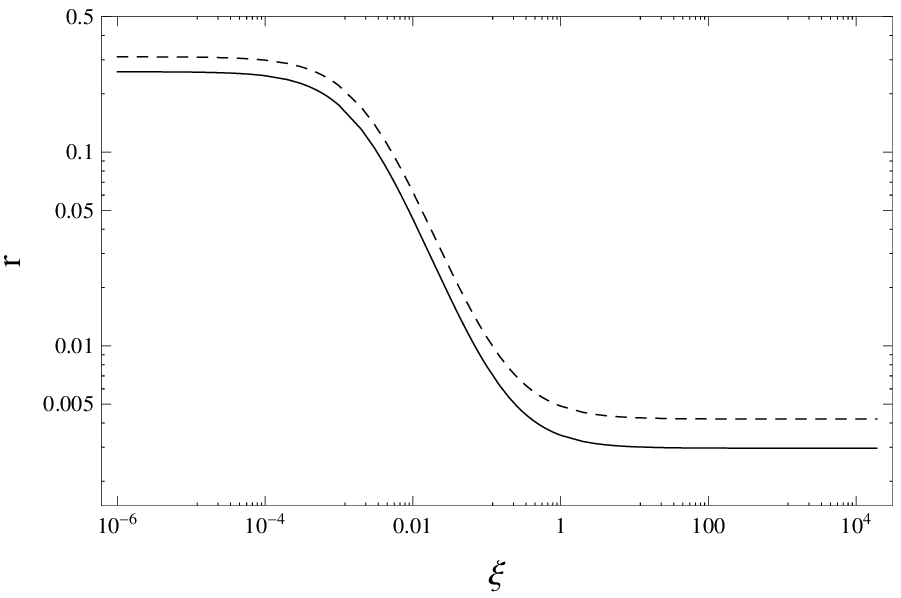}      \; 
\includegraphics[scale =0.85]{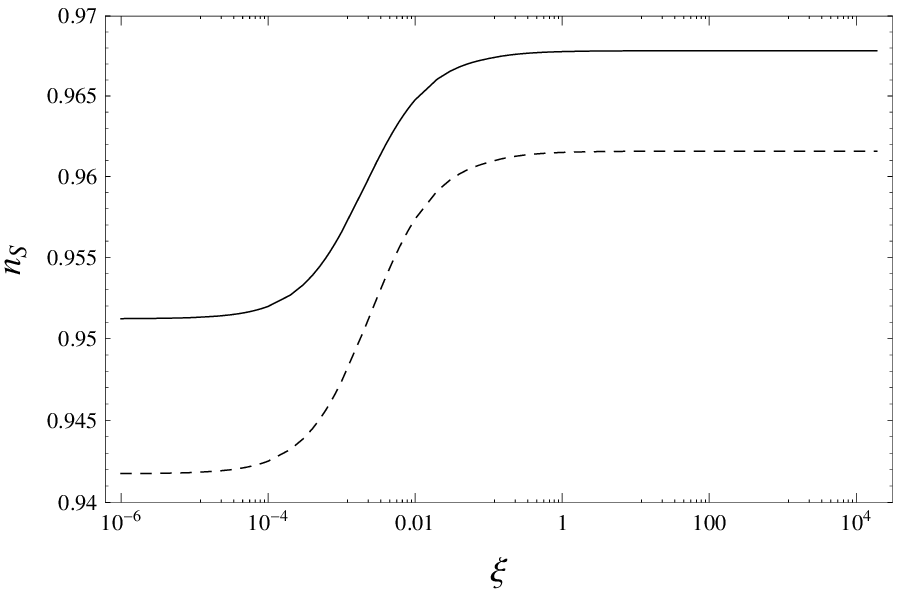} 
\includegraphics[scale=0.85]{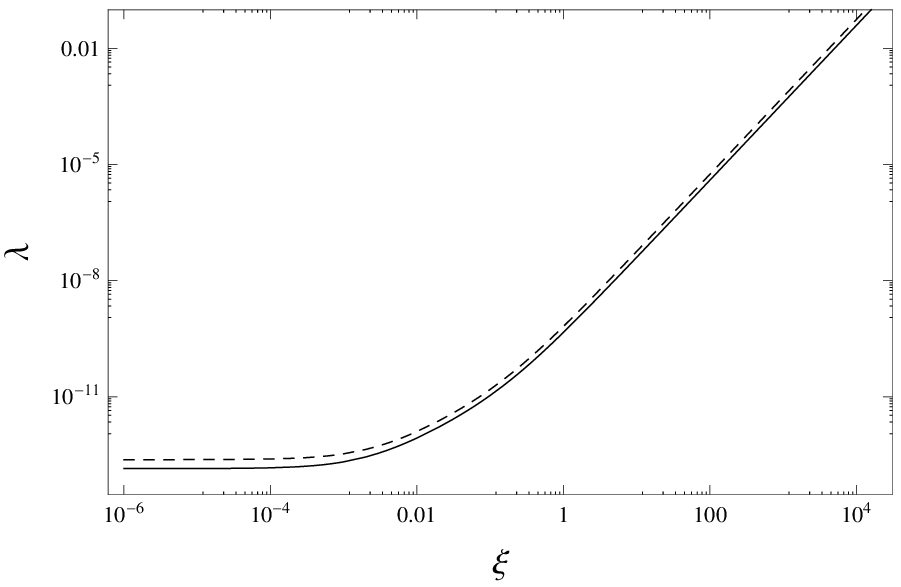}       \; 
\includegraphics[scale =0.85]{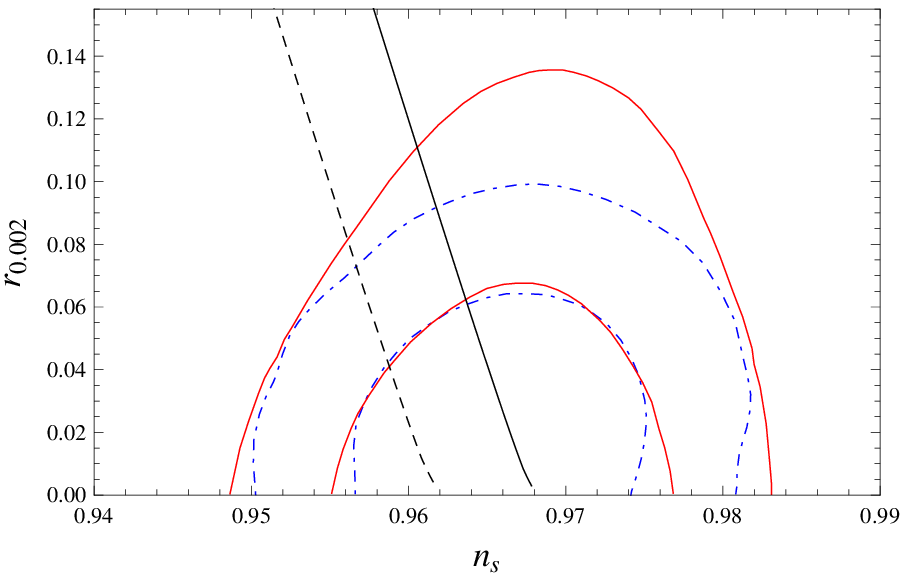}
\end{center}
\caption{
The inflationary predictions for $N=50$ (dot-dashed) and $N=60$ (solid) 
   in the non-minimal $\lambda \phi^4$ inflation.  
The top panels show $r$ vs. $\xi$ (left) and $n_s$ vs. $\xi$ (right). 
The bottom-left panel shows $\lambda$ vs. $\xi$. 
The inflationary predictions $n_s$ and $r$ for various values of $\xi$ are depicted in  the bottom-right panel,  
   along with the contours at the confidence levels of 68\% and 95\% 
   given by the results of Planck 2015 (solid) and Planck+BICEP2/Keck Array (dot-dashed) \cite{Planck2015}. 
}
\label{fig1}
\end{figure}

The inflationary slow-roll parameters in terms of the original 
 scalar field ($\phi$) are expressed as 
\bea
\epsilon(\phi)&=&\frac{1}{2} \left(\frac{V_E'}{V_E \sigma'}\right)^2, 
 \nonumber \\
\eta(\phi)&=& 
\frac{V_E''}{V_E (\sigma')^2}- \frac{V_E'\sigma''}{V_E (\sigma')^3} \ ,  
 \nonumber \\
\zeta (\phi) &=&  \left(\frac{V_E'}{V_E \sigma'}\right) 
 \left( \frac{V_E'''}{V_E (\sigma')^3}
-3 \frac{V_E'' \sigma''}{V_E (\sigma')^4} 
+ 3 \frac{V_E' (\sigma'')^2}{V_E (\sigma')^5} 
- \frac{V_E' \sigma'''}{V_E (\sigma')^4} \right)  ,
 \label{SRp}
\eea
where $V_E$ is the potential in the Einstein frame in terms of original field $\phi$ given by
\bea
V_E(\sigma(\phi)) = \frac14 \frac{\lambda \  \phi^4}{(1 +\xi \phi^2)^2} \ .
\eea

The amplitude of the curvature perturbation $\Delta_{\mathcal{R}}$ is given by 
\begin{equation} 
\Delta_{\mathcal{R}}^2 = \left. \frac{V_E}{24 \pi^2 \epsilon } \right|_{k_0},
\end{equation}
  which should satisfy $\Delta_\mathcal{R}^2= 2.195 \times10^{-9}$
  from the Planck 2015 results \cite{Planck2015} with the pivot scale chosen at $k_0 = 0.002$ Mpc$^{-1}$.
The number of e-folds is given by
\begin{eqnarray}
  N = \frac{1}{2} \int_{\phi_{\rm e}}^{\phi_I}
\frac{d\phi}{\sqrt{\epsilon(\phi)}}\left(\frac{d\sigma}{d\phi}\right)  ,
\end{eqnarray} 
where $\phi_I$ is the inflaton value at horizon exit corresponding to the scale $k_0$, 
  and $\phi_e$ is the inflaton value at the end of inflation, 
  which is defined by ${\rm max}[\epsilon(\phi_e), | \eta(\phi_e)| ]=1$.
The value of $N$ depends logarithmically on the energy scale during inflation 
  as well as on the reheating temperature, and is typically around 50--60.

The slow-roll approximation is valid as long as the conditions 
   $\epsilon \ll 1$, $|\eta| \ll 1$ and $\zeta \ll 1$ hold. 
In this case, the inflationary predictions, 
   the scalar spectral index $n_{s}$, the tensor-to-scalar ratio $r$, 
   and the running of the spectral index $\alpha=\frac{d n_{s}}{d \ln k}$, are given by
\bea
n_s = 1-6\epsilon+2\eta, \; \; 
r = 16 \epsilon,  \; \;
\alpha=16 \epsilon \eta - 24 \epsilon^2 - 2 \zeta. 
 \label{IP}
\eea 
Here the slow-roll parameters are evaluated at $\phi=\phi_I$.

Fig.~\ref{fig1} shows inflationary predictions for the non-minimal $\lambda \phi^4$ inflation 
  at the tree-level for $N=50$ (dashed-dotted) and $N=60$ (solid). 
Top panels show $r$ vs. $\xi$ (left) and $n_s$ vs. $\xi$ (right). 
Both $r$ and $n_s$ show asymptotic behavior for both small and large $\xi$ values. 
In the minimal $\lambda \phi^4$ inflation limit with $\xi = 0$, we obtain 
   $r \simeq 0.31$ $(0.26)$ and $n_s \simeq 0.942$ $(0.951)$ for $N=50$ $(60)$. 
The plots also show that for a larger e-holding number, we obtain a larger $n_s$ while a smaller $r$. 
The bottom-left panel shows the tree level quartic coupling $\lambda$ as a function of $\xi$. 
Note that $\lambda$ is very small unless $\xi \gg 1$. 
The inflationary predictions for $n_s$ and $r$ for various values of $\xi$ are depicted 
  in the bottom-right panel along with the results from the measurements by Planck 2015 and 
  Planck+BICEP2/Keck Array \cite{Planck2015}. 
We see that the inflationary predictions for $\xi \gtrsim 0.001$ are consistent with the observations.

\section{Running $B-L$ Higgs inflation and stability of inflaton potential}

\begin{table}[h]
\begin{center}
\begin{tabular}{c|ccc|c}
            & SU(3)$_c$ & SU(2)$_L$ & U(1)$_Y$ & U(1)$_{B-L}$  \\
\hline
$ q_L^i $    & {\bf 3}   & {\bf 2}& $+1/6$ & $+1/3$  \\ 
$ u_R^i $    & {\bf 3} & {\bf 1}& $+2/3$ & $+1/3$  \\ 
$ d_R^i $    & {\bf 3} & {\bf 1}& $-1/3$ & $+1/3$  \\ 
\hline
$ \ell^i_L$    & {\bf 1} & {\bf 2}& $-1/2$ & $-1$  \\ 
$ N\!R^i$   & {\bf 1} & {\bf 1}& $ 0$   & $-1$  \\ 
$ e_R^i  $   & {\bf 1} & {\bf 1}& $-1$   & $-1$  \\ 
\hline 
$ H$         & {\bf 1} & {\bf 2}& $-1/2$  &  $ 0$  \\ 
$ \varphi$      & {\bf 1} & {\bf 1}& $  0$  &  $+2$  \\ 
\end{tabular}
\end{center}
\caption{
Particle contents of the minimal $B-L$ model. 
In addition to the SM particle contents, the right-handed neutrino $N\!R^i$ 
 ($i=1,2,3$ denotes the generation index)  and a complex scalar $\varphi$ are introduced. 
}
\end{table}

In order to investigate the Higgs inflation with the stabilized inflaton potential,  
   in this paper we take the minimal $B-L$ extension of the SM as an example,  
   where the anomaly-free U(1)$_{B-L}$ gauge symmetry is introduced 
   along with a scaler field $\varphi$ and three right-handed neutrinos $N\!R^i$. 
The particle contents of our model are listed in Table~1. 
This model requires three generations of right-handed neutrinos to cancel all the gauge and gravitational anomalies. 
The $B-L$ gauge symmetry is broken by the vacuum expectation value (VEV) of $\varphi$ 
   in its Higgs potential of 
\bea
  V(|\varphi|) = \lambda \left( \varphi^\dagger \varphi - \frac{v_{BL}}{2}  \right)^2.  
\eea
Associated with the gauge symmetry breaking, 
    the right-handed neutrinos acquire their Majorana masses through the Yukawa interaction,    
\bea  
   {\cal L} \supset  - \frac{1}{2} \sum_{i=1}^{3} Y  \varphi  \overline{N\!R^{i c}} N\!R^i  +{\rm h.c.},
\eea
where we have taken the degenerate mass spectrum for the right-handed neutrinos, for simplicity.  
After the $B-L$ symmetry breaking with the Higgs VEV $\langle \varphi \rangle=v_{BL}/\sqrt{2} $,  
   the particle masses are given by 
\bea 
m_{Z^\prime}= 2 \; g \; v_{BL},  \; \; 
m_{NR}^i= \frac{1}{\sqrt{2}} \; Y \; v_{BL}, \; \; 
m_{\phi} = \sqrt{2 \lambda} \;  v_{BL}. 
\label{masses}
\eea

Let us now consider the $B-L$ Higgs inflation scenario. 
The action in the Jordan frame is given by 
\bea
\mathcal{S}_J^{tree} = \int \mathrm{d}^4 x  \sqrt{-g} 
   \left[ -\frac{1}2f(|\varphi|) \mathcal{R} +(D_{\mu} \varphi)^{\dagger}g^{\mu \nu}(D_{\nu} \varphi) 
   - V(|\varphi|)  \right], 
\eea
where $D_{\mu}= \partial_\mu - i 2 g Z^\prime_\mu$, and 
   $f(|\varphi|) \equiv 1 +2 \xi \varphi^{\dagger} \varphi$, and 
   $\varphi= (v_{BL} + \phi)/\sqrt{2}$ in the unitary gauge with the physical Higgs filed $\phi$ 
   identified as inflaton. 
In the Einstein frame, the RG improved effective inflaton potential at the one-loop level 
    is given by \cite{George:2013/15}
\bea 
  V_E(\phi) = \frac{1}{4} \lambda(\Phi) \; \Phi^4 ,
\end{eqnarray}
   where $\Phi \equiv \phi/\sqrt{1 +\xi \phi^2} $, and we have neglected $v_{BL}$ 
   which is much smaller than the Planck mass.   
The RG equations of the couplings at the one-loop level are given by \cite{ORS-BL}
\bea
16 \pi^2  \mu  \frac{d g}{d \mu} &=& \frac{1}{3} (4s+32) g^3,         \nonumber\\
16 \pi^2  \mu \frac{d Y}{d \mu}   &=& -6 g^2 Y+\frac{5}{2} Y^3 ,   \nonumber\\
16 \pi^2  \mu \frac{d \lambda}{d \mu}  &=& \left(18 s^2+2\right) \lambda^2- (48 g^2 -6 Y^2) \lambda +96 g^4 - 3 Y^4 , 
  \label{RGEs}
\eea
where a factor $s$ defined as 
\begin{equation}
  s =  \frac{(1 +\xi \mu^2)}{1+(1+6\xi)\xi \mu^2} 
\end{equation}
  is assigned to each term in the RG equations associated with only the physical Higgs boson 
  loop corrections \cite{Higgs_inflation1, Higgs_inflation2}.\footnote{
There are a few different prescriptions for computing quantum corrections  
   in the presence of the non-minimal gravitational coupling \cite{Higgs_inflation3}. 
For recent, detailed computations of quantum corrections,    
  see \cite{George:2013/15}  and their results of 1-loop beta functions with the $s$-factor. 
} 
We find that this $s$-factor has no effect in our numerical analysis 
  for a parameter region satisfying $g^2, Y^2 \gg \lambda$, 
  and one may fix $s=1$ as a good approximation.

\begin{figure}[t]
\begin{center}
\includegraphics[scale=0.88]{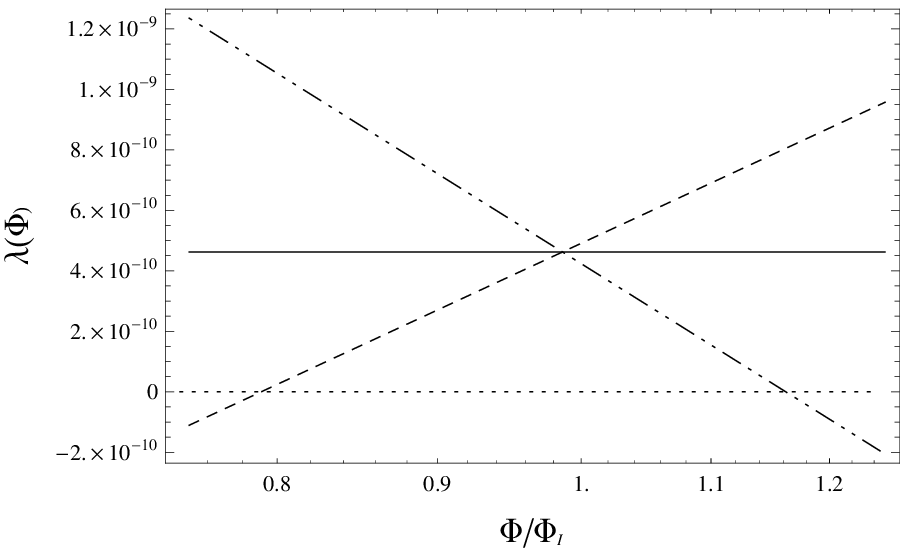}  \;  
\includegraphics[scale =0.88]{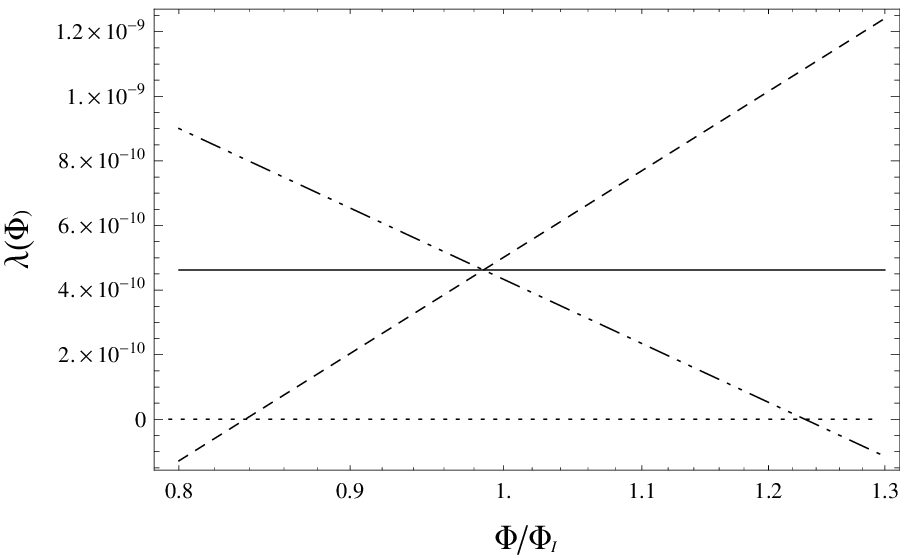}
\end{center}
\caption{
The RG evolution of the inflaton quartic coupling $\lambda$ for various values of 
  $g$ and $Y$, with the fixed value of $\xi =1$.   
The initial value of $\lambda(\Phi_I)$ is fixed by the tree-level analysis. 
In the left panel, the solid, the dashed and the dot-dashed lines show 
   the running quartic couplings for $g=0.01$, $0.011$ and $0.009$, respectively, 
   with the fixed value of  $Y=0.0237$. 
In the right panel, the solid, the dashed and the dot-dashed lines show 
   the running couplings for $Y=0.0237$, $0.0214$ and $0.0261$, respectively, 
   with the fixed value of  $g=0.01$. 
The stability condition of $\beta_\lambda(\Phi_I)=0$ is satisfied for $g=0.01$ and $Y=0.0237$, 
   and the corresponding running couplings are depicted as the solid lines.   
}
\label{fig2} 
\end{figure}

Let us now investigate the stability of the effective inflaton potential. 
In our analysis throughout this paper, we set the initial values of $\lambda$  
   to be the one obtained in the tree-level analysis at the initial inflaton value $\phi=\phi_I$, 
   equivalently, $\Phi_I= \phi_I/\sqrt{1+\xi \phi_I^2} $. 
Then, we consider the RG improved effective inflaton potential by taking into account 
   the RG evolution of the quartic coupling with the initial condition at $\phi_I$.  
As we have seen in the previous section, the inflaton quartic coupling is very small unless $\xi \gg 1$. 
Hence the beta function of the quartic coupling is approximately given by 
\bea 
\beta_\lambda = 
\frac{1}{16 \pi^2}
   \left[ 
\left(18 s^2+2\right) \lambda^2- (48 g^2 -6 Y^2) \lambda +96 g^4 - 3 Y^4  
  \right] 
\simeq 
\frac{1}{16 \pi^2}  \left( 96 g^4 - 3 Y^4    \right) ,  
  \label{beta} 
\eea
  when $g^2, Y^2 \gg \lambda$.\footnote{
In this paper, we are interested in this case, otherwise the beta function is 
  so small that the inflaton quartic coupling is almost RG invariant.  
Although the tree-level analysis is valid in this case, 
  the gauge and Yukawa couplings are too small to yield any impacts 
  in the experimental point of view.
}   
The RG evolution is controlled by $g$ and $Y$, which are independent of 
   the initial value of the inflaton quartic coupling.  
Fig.~\ref{fig2} shows the RG evolution of the inflaton quartic coupling in the vicinity of 
   the initial inflaton value for various values of $g$ and $Y$ with a fixed $\xi=1$. 
In the left panel, the solid, the dashed and the dot-dashed lines denote the RG evolutions 
   for $g=0.01$, $0.011$ and $0.009$, respectively, with the fixed value of  $Y=0.0237$. 
In the right panel, the solid, the dashed and the dot-dashed lines denote the RG evolutions 
   for $Y=0.0237$, $0.0214$ and $0.0261$, respectively, with the fixed value of  $g=0.01$. 
Note that $\beta_\lambda(\Phi_I)=0$  is satisfied 
    with the parameter choice for the solid lines, $g=0.01$ and $Y=0.0237$.       
We can see from Fig.~\ref{fig2} that if the condition of $\beta_\lambda=0$ is violated 
   even with $\pm 10$ \% deviations for the values of $g$ or $Y$,  
   the running quartic coupling quickly becomes negative in the vicinity of $\Phi_I$  
   (see the dashed and the dot-dashed lines).  
This fact indicates that the $B-L$ gauge symmetry breaking vacuum at $\phi=v_{BL}$ is unstable 
   and the effective potential develops a true vacuum with a negative cosmological constant.  
The quantum corrections through the gauge and Yukawa coupling completely 
   change our inflationary scenario from the one at the tree-level.

\begin{figure}[h]
\begin{center}
\includegraphics[scale=0.89]{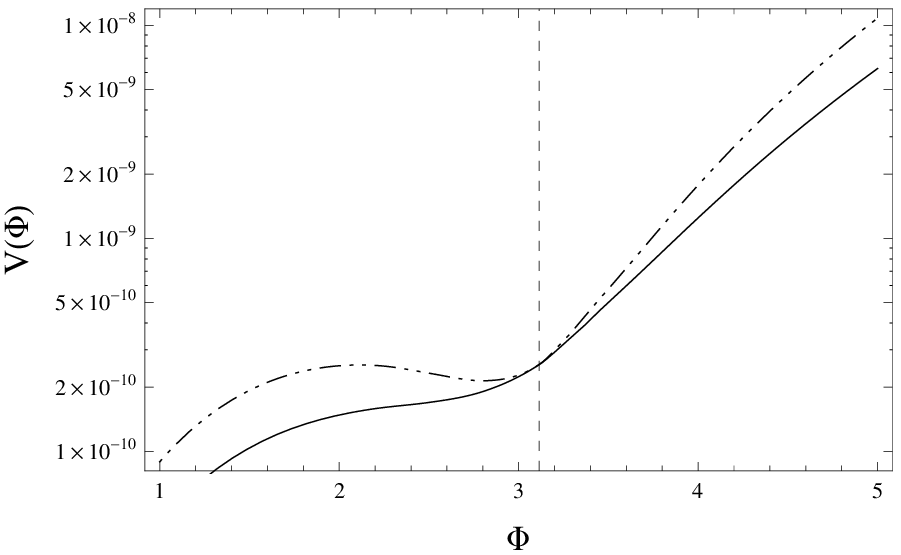}   \; 
\includegraphics[scale =0.87]{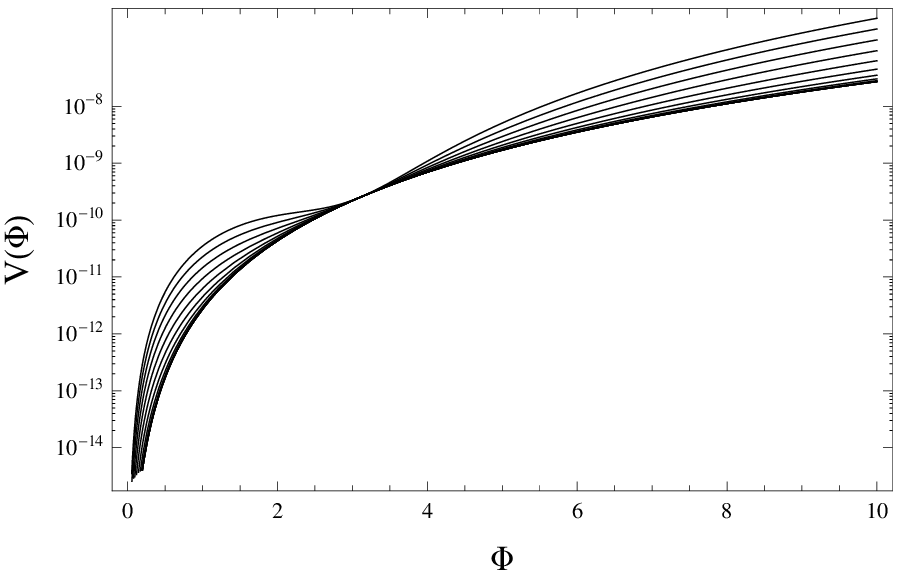}
\end{center}
\caption{
The RG improved effective potential for various values of $g$ and a fixed $\xi=0.1$.  
In the left panel, the effective potentials for $g = 0.041$ (solid) and $0.046$ (dot-dashed) are shown.  
For $g =0.046 > g_{max}=0.0425$, the local minimal is developed.  
The vertical dashed line indicates $\Phi=\Phi_I$ (in the Planck unit).   
In the right panel, the effective potentials for various values of $g_{min} <g < g_{max}$ are shown. 
 }
 \label{fig3}
\end{figure}

In order to avoid this instability, we impose not only the condition of $\beta_\lambda=0$ 
    but also $d \beta_\lambda/d \Phi > 0$  at $\Phi_I$. 
From the first condition, the Yukawa coupling $Y$ is determined by 
    the gauge coupling, which we take as a free parameter in our analysis, 
    along with the others, $\xi $  and $v_{BL}$.  
The second condition ensures that the effective potential is monotonically increasing 
    in the vicinity of $\Phi_I$, and yields a lower bound on $g > g_{min}$. 
When we analyze the global structure of the effective potential, we can notice  
    that there exists an upper bound on $g< g_{max}$.  
For a large $g > g_{max}$, the effective potential develops a local minimum at $\Phi < \Phi_I$,
    so that the inflaton field will be trapped in this minimum after inflation. 
A second inflation then takes place until the vacuum transition from this local minimum 
    to the true $B-L$ symmetry breaking vacuum. 
To avoid this problem, the parameter region is restricted to be in the range of $g_{min} < g < g_{max}$. 
Fig.~\ref{fig3} shows the effective potential for various values of the gauge coupling ($g$) for fixed $\xi =0.1$. 
In the left panel, the solid line depicts the effective potential for $g=0.041$,  
  while the dot-dashed line is for $g=0.046$.  
In this example, we find the upper bound as $g_{max}=0.0425$.  
We can see that the effective potential develops a local minimum for $g=0.046 > g_{max}$. 
For various values of $g < g_{max}$, the effective potentials are shown in the right panel.

\section{Inflationary predictions and low energy observables}
Under the stability conditions, $g_{min} < g < g_{max}$ with various value of $\xi$, 
   we now calculate the inflationary predictions with the effective inflaton potential. 
Since we refer the results in the tree-level analysis for $\lambda(\Phi_I)$ for a fixed $\xi$ 
   and impose the stability condition $\beta_\lambda (\Phi_I) =0$,   
   our prediction for the tensor-to-scalar ratio $r$ is the same as the one obtained in the tree-level analysis. 
However, the RG evolution of the inflaton quartic coupling alters the other inflationary predictions, 
   $n_s$ and $\alpha$, from those obtained in the tree-level analysis,  
   because they are calculated by the second and third derivatives of the effective potential 
   (see Eqs.~(\ref{SRp}) and(\ref{IP})).

Let us first derive a analytic formula for the deviations of $n_s$ from the tree-level prediction. 
In the effective inflaton potential, the inflation quartic coupling is not a constant, but a function of $\Phi=\phi/\sqrt{1+\xi \phi^2}$. 
With the stability condition $\beta_\lambda (\Phi_I)=0$,  
    we calculate the derivatives of the effective potential as 
\bea 
    V_E^\prime(\phi_I) &=&  \frac{1}{4} \lambda^\prime (\Phi) \;  \Phi^4 \Big|_{\phi=\phi_I} 
                                   +  \frac{1}{4} \lambda (\Phi) \left( \Phi^4 \right)^\prime  \Big|_{\phi=\phi_I}
                                   =  \frac{1}{4} \lambda (\Phi) \left( \Phi^4 \right)^\prime  \Big|_{\phi=\phi_I}  
                                   =  V_E^\prime \Big|_{tree}   ,
\label{1stD}  \\ 
  V_E^{\prime \prime}(\phi_I) &=&  \frac{1}{4} \lambda^{\prime \prime} (\Phi) \;  \Phi^4 \Big|_{\phi=\phi_I} 
                                   +  \frac{1}{2} \lambda^\prime (\Phi) \left( \Phi^4 \right)^\prime  \Big|_{\phi=\phi_I}
                                   +  \frac{1}{4} \lambda (\Phi) \left( \Phi^4 \right)^{\prime \prime}  \Big|_{\phi=\phi_I}  
  \nonumber\\ 
    &=&  \frac{1}{4} \lambda^{\prime \prime} (\Phi) \;  \Phi^4 \Big|_{\phi=\phi_I} 
            + V_E^{\prime \prime} \Big|_{tree},      
\label{2ndD}
\eea
where a prime denotes the derivative with respect to $\phi$,
     $V_E^\prime \Big|_{tree} $ and $V_E^{\prime \prime} \Big|_{tree}$ are evaluated 
     to be the same as those in the tree-level analysis, 
     and we have used $\lambda^\prime(\phi_I) \propto \beta_\lambda(\Phi_I)=0$ under the stability condition. 
Rewriting $d/d\phi$ in terms of $d/d \Phi$, we have 
\bea
    \epsilon(\phi_I)  = \epsilon_{tree},  \; \; \;  \eta(\phi_I) = \eta_{tree} + \Delta \eta ,
\eea
where $\epsilon_{tree}$ and $\eta_{tree}$ are the slow-roll parameters evaluated in the tree-level analysis, and   
\bea
   \Delta \eta = \frac{(1-\xi \Phi_I^2)^2}{1+6 \xi^2 \Phi_I^2}  \; 
       \frac{\ddot{\lambda}(\Phi_I)}{\lambda (\Phi_I)}  
    =  \frac{(1-\xi \Phi_I^2)^2}{1+6 \xi^2 \Phi_I^2}  \; 
       \frac{{\dot \beta}_\lambda (\Phi_I)}{\Phi_I \lambda (\Phi_I)}   .
\eea
Here, a dot is the derivative with respect to $\Phi$.  
Since ${\dot \beta}_\lambda$ is non-zero, $\eta$ is deviated from its tree-lvel value $\eta_{tree}$. 
Using Eqs.~(\ref{RGEs}) and (\ref{beta}), we obtain
\bea 
   16 \pi^2 \Phi_I {\dot \beta}_\lambda (\Phi_I) \simeq  
     4608 \; g^6  +72 \; g^2 Y^4  -30 \; Y^6   \simeq 768  \left(  9 -5 \sqrt{2} \right) g^6 , 
\eea
where in the last expression, we have used the stability condition $Y^4 \simeq  32 g^4$ at $\Phi_I$. 
Finally, we arrive at an expression for the spectral index as (see Eq.~(\ref{IP}))\footnote{
In the same way, we can express the running of the spectral index, $\alpha$,  
   in terms of $\xi$, $\Phi_I$, $\lambda$ and $g$.  
However, as we will see in the following numeral analysis, 
   the predicted $\alpha$ values are found to be very small and always consistent with the Planck 2015 results. 
Thus, we omit the expression for $\alpha$. 
}
\bea 
  n_s = n_s^{tree} + 2 \Delta \eta \simeq n_s^{tree} + 
       \frac{6 (9-5 \sqrt{2})}{\pi^4}  \; 
        \frac{(1-\xi \Phi_I^2)^2}{(1+6 \xi^2 \Phi_I^2) \Phi_I^2}  \;  \frac{g(\Phi_I)^6}{\lambda(\Phi_I)}  .
 \label{deviations}
\eea
For a fixed $\xi$, $\Phi_I$ and $\lambda(\Phi_I)$ are determined by the tree-level analysis, 
   so that the inflationary prediction is controlled by the gauge coupling $g$, 
   which is the free parameter in our analysis.

In our model, there are only three free parameters, $\xi$, $v_{BL}$ and $g$, 
   with a fixed e-folding number $N=50/60$.  
Once we fix $\xi$ and $N$, 
   $\Phi_I$ and $\lambda(\Phi_I)$ are fixed by the tree-level analysis for inflation, 
   and the inflationary predictions except for $r$ are controlled by the gauge coupling $g$ 
   with its relation to the Yukawa coupling $Y$ led by the stability condition $\beta_\lambda(\Phi_I)=0$. 
In the $B-L$ model, the particle mass spectrum is determined by the gauge, Yukawa and inflation quartic couplings 
   at the scale $v_{BL}$ (see Eq.~(\ref{masses})), 
   which are obtained by solving the RG equations in Eq.~(\ref{RGEs}) from $\mu=\Phi_I$ to $\mu=v_{BL}$. 
Since $\lambda(\Phi_I)$ is very small, its RG evolution is determined by $g$ and $Y$ 
   as we can see from its RG equation. 
Considering all of these facts, we expect that there exist a nontrivial mass relation in the particle spectrum 
   and a nontrivial correlation between the inflationary predictions and the particle mass spectrum. 
In the following, the results of our numerical analysis will show such non-trivial relations.   
For simplicity, we fix $v_{BL}$ so as to yield $m_{Z^\prime}= 2 g(v_{BL}) v_{BL}=3$ TeV, 
   to be consistent with the current results from the search for $Z^\prime$ boson resonance 
   at the Large Hadron Collider \cite{ATLAS_Zp, CMS_Zp}.

\begin{figure}[h]
\begin{center}
\includegraphics[scale=1]{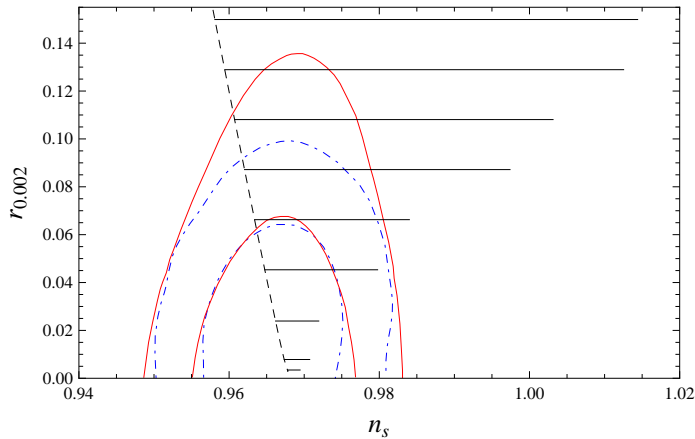} 
\end{center}
\caption{
Inflationary predictions for various values of $\xi$ with the inputs of $g$ varied in the range of 
   $g_{min} < g < g_{max}$, 
   along with the contours given by the experiments (same as in Fig.~\ref{fig1}). 
The horizontal solid lines from top to bottom 
   correspond to the results for $\xi= 1.5 \times 10^{-3}$, $2.1 \times 10^{-3}$, $2.9 \times 10^{-3}$, 
    $4.1 \times 10^{-3}$, $6 \times \!10^{-3}$, $0.01$, $0.02$,  $0.1$ and $1$.
The diagonal dashed line denotes the inflationary predictions in the tree-level analysis.
\label{fig4}
}
\end{figure}

Fig.~\ref{fig4} shows the resultant inflationary predictions for a variety of $\xi$ values 
   with the input values of $g$ in the range of $g_{min} < g < g_{max}$ at $\Phi_I$ ($N=60$), 
   along with the contours given by the Planck 2015 results.  
As we have discussed above, the prediction for the tensor-to-scalar ratio is the same 
   as the one in the tree-level analysis, while the predicted spectral index is altered 
   by quantum corrections. 
In Fig.~\ref{fig4}, we can see that for $\xi \lesssim 1$, the results show sizable deviations 
   for $g \sim g_{max}$ from those at the tree-level analysis depicted as the diagonal dashed line. 
Interestingly, the Planck 2015 results provide upper bounds on $g$,  
   which are more severe than $g_{max}$ for $\xi \lesssim 0.001$.  
In our numerical analysis, we can see that $\Phi_I$ approaches $1/\sqrt{\xi}$ from a smaller value 
   as we increase $\xi \gtrsim 1$, and hence the deviation of the predicted $n_s$ value 
   from the one in the tree-level analysis becomes smaller 
   as we can see from Eq.~(\ref{deviations}) with the limit $1-\xi^2 \Phi_I \to 0$.

\begin{figure}[h]
\begin{center}
\includegraphics[scale=0.83]{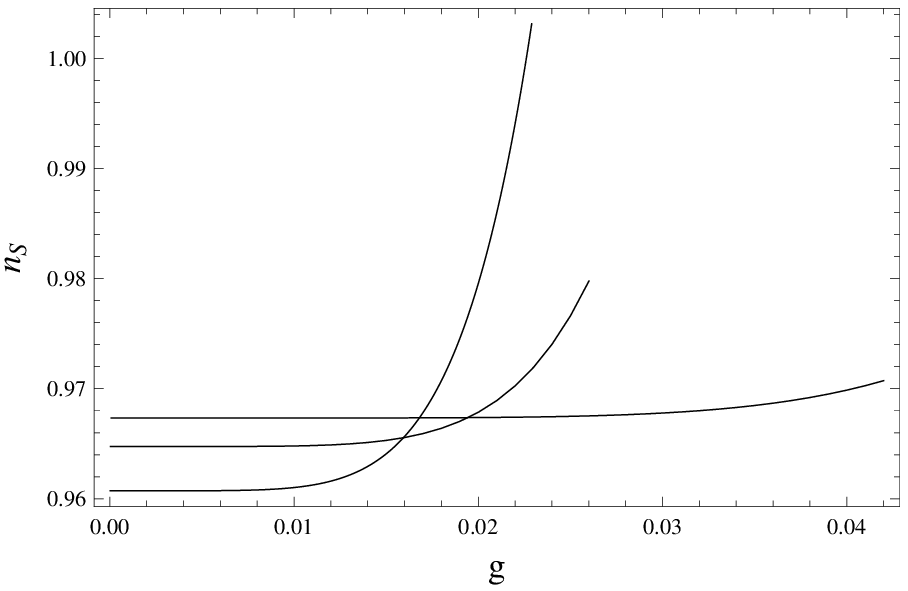}    \; 
\includegraphics[scale =0.88]{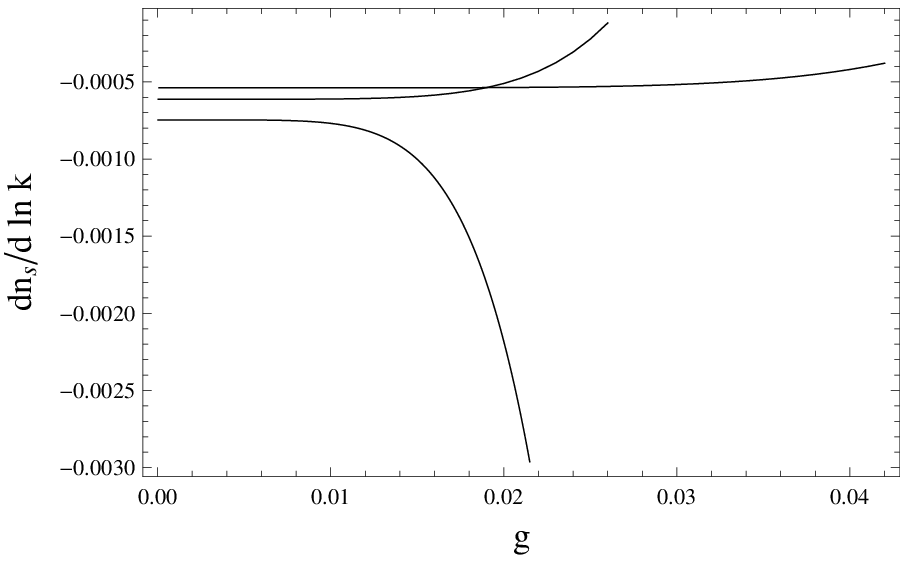}
\includegraphics[scale =0.88]{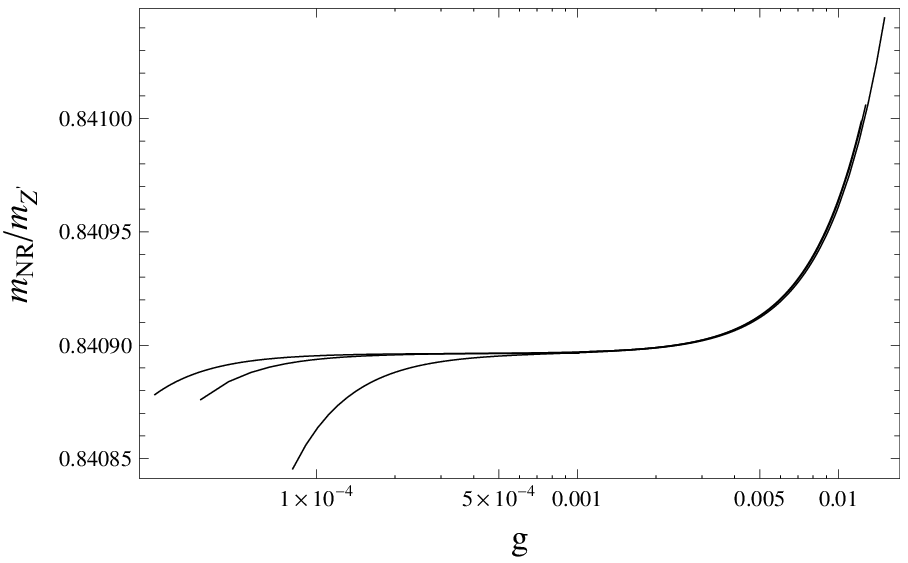} \; 
\includegraphics[scale=0.83]{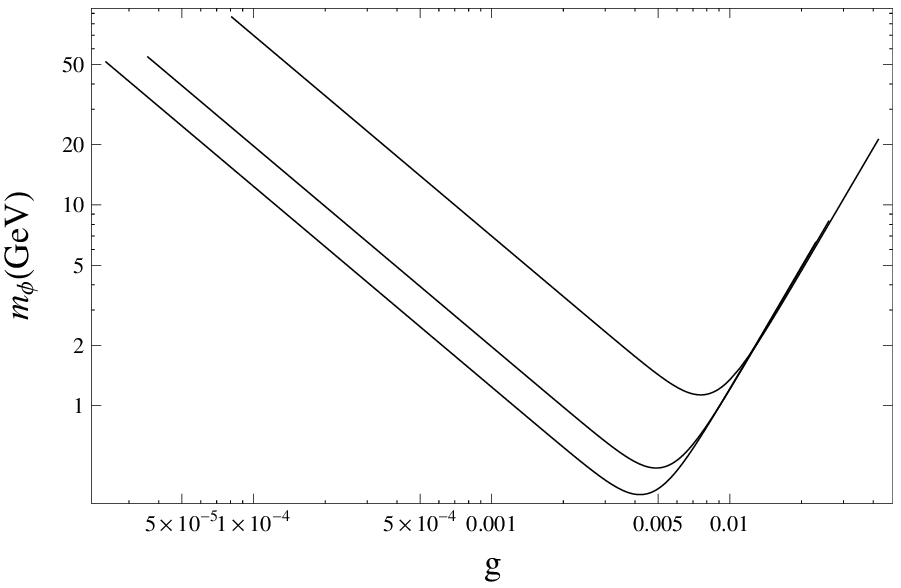}    
\end{center}
\caption{
Inflationary predictions and low energy observables 
  for various values of $g < g_{max}$ and $\xi \simeq 0.0029, 0.01$ and $0.1$ 
 (corresponding $r \simeq 0.108, 0.045$ and $0.008$), respectively. 
We have fixed $N=60$ and $m_{Z^\prime}$ to be $3$ TeV.
$g_{max}$ increases with increasing $\xi$. 
Top two panels show the inflationary predictions for spectral index $n_s$ (left) and 
   the running of spectral index $d n_s / d \ln k$ (right), for decreasing $r$ (top to bottom). 
Bottom left panel shows mass of the inflaton  $m_\phi$ (top to bottom) for decreasing $r$. 
Bottom right panel shows  mass ratio $m_{NR} / m_{Z^\prime}$ (top to bottom) for increasing $r$ values. 
}
\label{fig5}
\end{figure}

In order to see the inflationary predictions as a function of $g$, 
    we show our results in Fig.~\ref{fig5} for $\xi=0.0029$, $0.01$ and $0.1$, with $N=60$. 
The top-left and top-right panels show the inflationary predictions of $n_s$ and $\alpha=dn_s/d \ln k$ 
    as a function of $g$ in the range of $g_{min} < g < g_{max}$.  
The prediction for $r$ is the same in the tree-level analysis, and  
   $r \simeq 0.108, 0.045$ and $0.008$, respectively, for $\xi=0.0029$, $0.01$ and $0.1$.      
For a larger value of $\xi$, $g_{max}$ becomes larger. 
As $g$ is lowered, the predicted $n_s$ value approaches the tree-level prediction. 
By numerically solving the RG equations for the couplings in Eq.~(\ref{RGEs}) for a fixed $g$, 
    we obtain the particle mass spectrum with $m_{Z^\prime}=3$ TeV.  
The mass ratio $m_N/m_{Z^\prime}$ is shown in the bottom-left panel, 
   while the bottom-right panel shows the inflaton mass. 
In both panels, the solid lines from left to right correspond to the results for 
   $\xi=0.0029$, $0.01$ and $0.1$, respectively. 
The resultant mass ratio is almost independent of $\xi$, but shows a splitting for $g \lesssim 5 \times 10^{-4}$. 
For such a very small $g$, its corresponding $Y$ determined by the stability condition is also very small, 
   and hence both $g$ and $Y$ are almost RG invariant and   
   the mass ratio is determined by $Y(\Phi_I)/g(\Phi_I)$. 
However, in this case, the condition $g^2, Y^2 \gg \lambda$ is no longer valid, 
   and $Y$ determined by $\beta_\lambda(\Phi_I)=0$ depends on the input $\lambda$ values. 
This is the reason why the bottom-left panel shows the splitting among three solid lines 
   for $g \lesssim 5 \times 10^{-4}$. 
For $g \gtrsim 0.005$,  the RG evolution of $\lambda$ is mainly determined by 
   $g$ and $Y$ in its beta function, since $\lambda(\Phi_I)$ is extremely small.  
For a very small $g$ value $\lesssim 0.005$, the effect of $g$ and $Y$ 
  on the RG evolution of $\lambda$ becomes negligible, 
  and $\lambda(v_{BL}) \simeq \lambda(\Phi_I)$.  
Since we have fixed $m_{Z^\prime}=2 g(v_{BL}) v_{BL}=3$ TeV,  
   $v_{BL} = 1.5 \;{\rm TeV}/g(v_{BL}) \simeq 1.5 \;{\rm TeV}/g(\Phi_I)$,  
   and the inflaton mass becomes larger proportionally to $1/g(\Phi_I)$ 
   as shown in the bottom-right panel.

\begin{figure}[h]
\begin{center}
\includegraphics[scale=0.8]{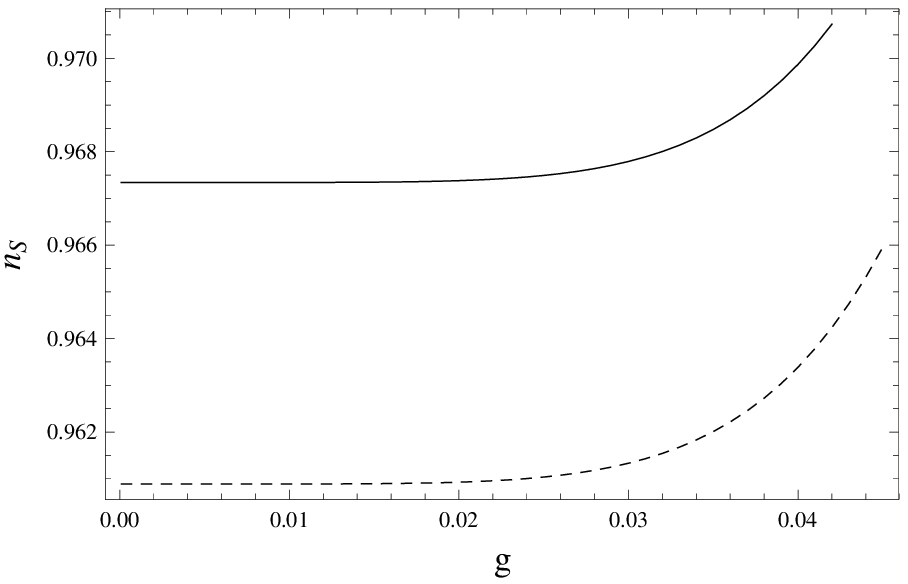}       \;
\includegraphics[scale =0.8]{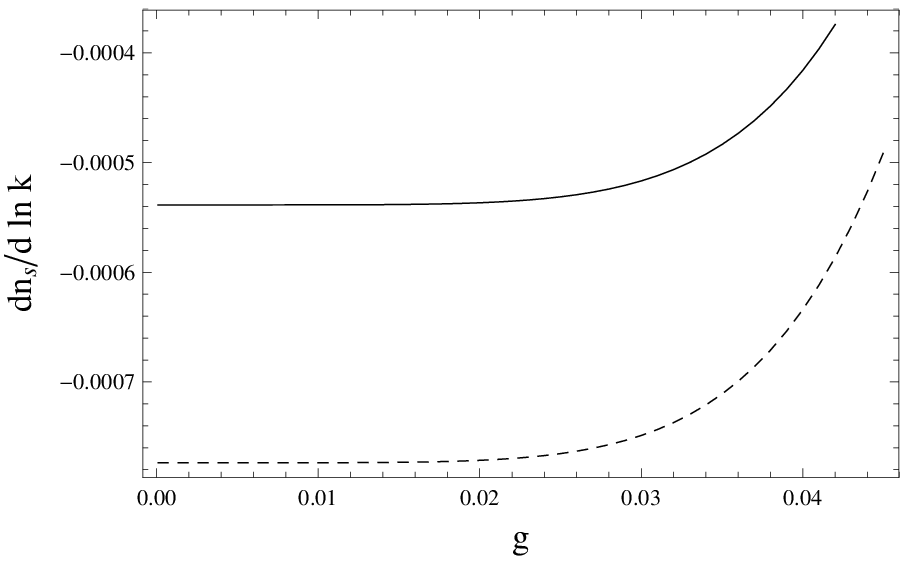}
\includegraphics[scale =0.8]{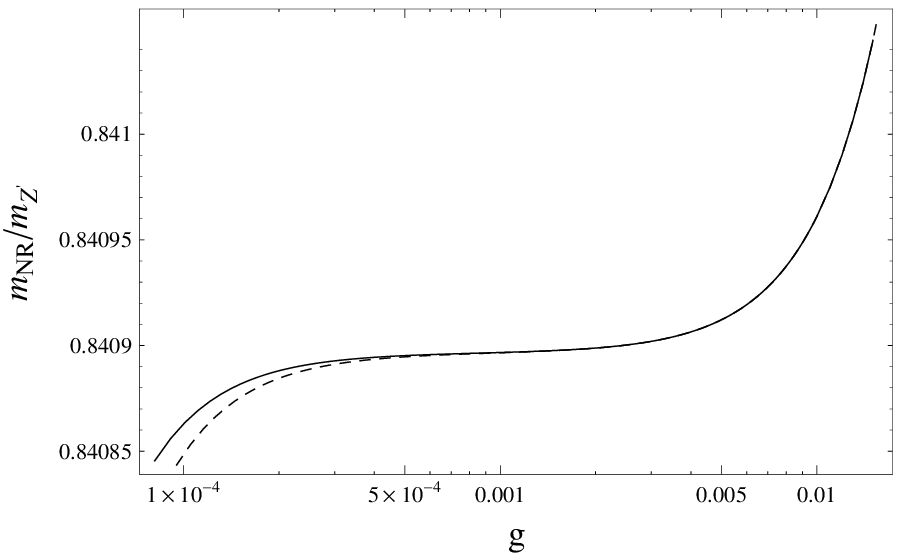}  \; 
\includegraphics[scale=0.8]{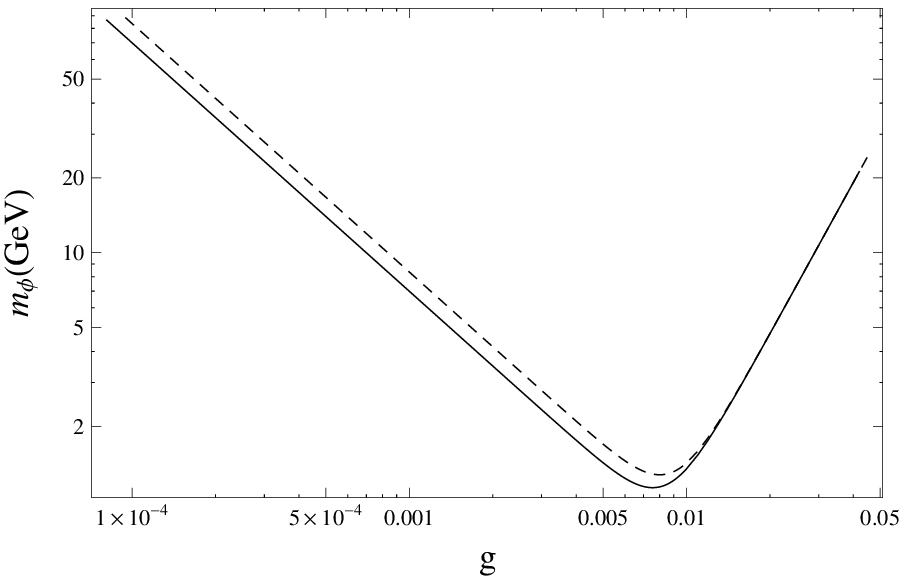}  
\end{center}
\caption{
Same as Fig.~\ref{fig5} with $\xi=0.1$, but for two different values of 
   $N=50$ (dashed lines) and $N=60$ (solid lines). 
}
\label{fig6}
\end{figure}

Same as Fig.~\ref{fig5} but for $N=50$ and $60$ with $\xi=0.1$ is depicted in Fig.~\ref{fig6}. 
The dashed lines denote the results for $N=50$,  while the solid lines for $N=60$. 
The inflationary predictions show a sizable deference for the two different $N$ values,  
   as shown in the tree-level analysis in Sec.~2. 
On the other hand, the particle mass spectrum weakly depends on $N$.

\begin{figure}[h]
\begin{center}
\includegraphics[scale=0.8]{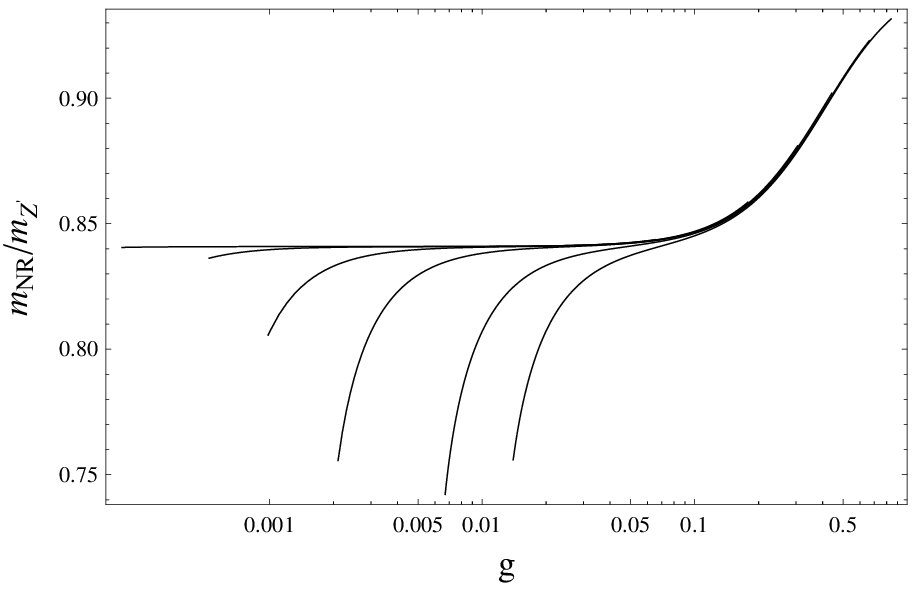}    \;
\includegraphics[scale=0.8]{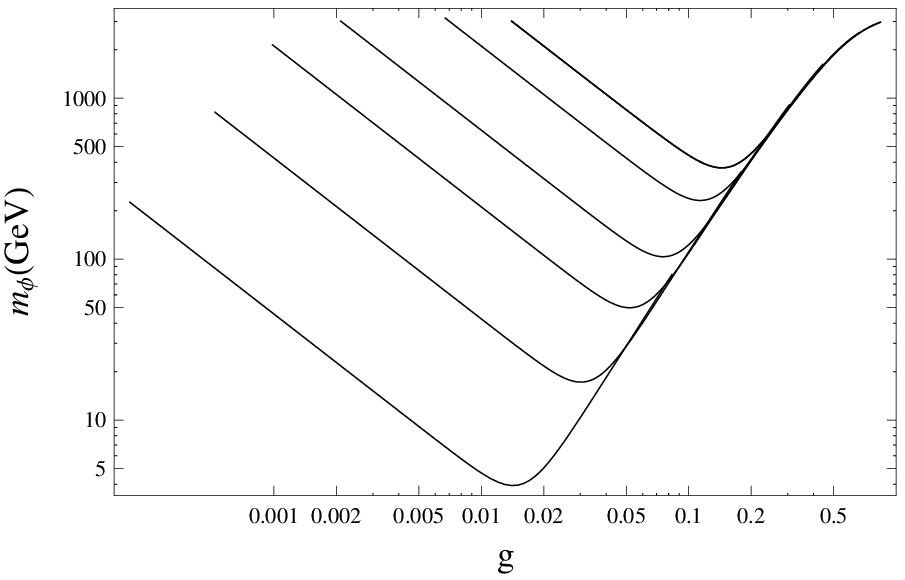} \; 
\end{center}
\caption{
Mass spectrum for various value of 
  $\xi =1, 10, 50, 150 ,500$ and $1000$ from left to right.  
The left panel shows the mass ratio $m_{NR} / m_{Z^\prime}$,  
  while the right panel shows the mass of the inflaton. 
Here we have fixed $m_{Z^\prime}=3$ TeV. 
}
\label{fig7}
\end{figure}

We show the results for the mass spectrum for large $\xi$ values and $N=60$ in Fig.~\ref{fig7} 
The solid lines from left to right corresponds to the results for $\xi =1$, $10$, $50$, $150$, $500$ and $1000$. 
For $g\lesssim0.05$, the resultant mass ratio in the left panel shows $\xi$-dependence. 
This is because $g^2, Y^2 \gg \lambda$ is no longer valid for such a small $g$ value,  
   and the $Y$ value determined by $\beta_\lambda(\Phi_I)$ depends on $\lambda(\Phi_I)$. 
We also show the inflation mass spectrum in the right panel, which show a similar behavior 
   as the results in the bottom-right panel in Fig.~{\ref{fig5}}.    
As we have seen in Fig.~\ref{fig4}, the inflationary predictions for $\xi \gtrsim 1$ are close to 
  those obtained in the tree-level analysis, $n_s \simeq 0.968$ and $r \simeq 0.003$.

\begin{figure}[h]
\begin{center}
\includegraphics[scale=0.8]{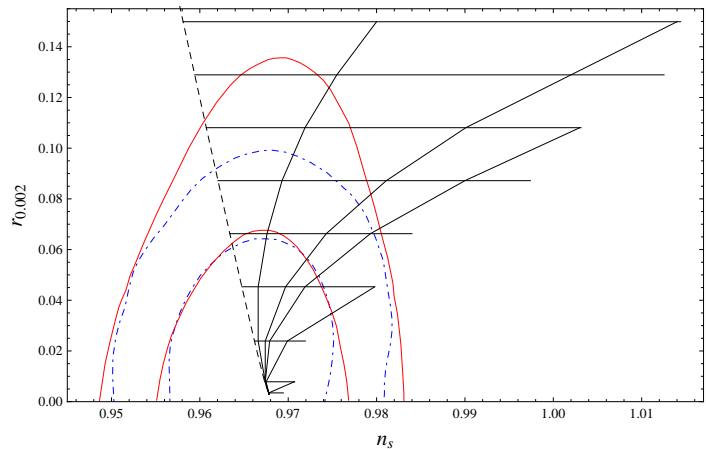}
\end{center}
\caption{
Inflationary predictions for various fixed $g(v_{BL})$ values along with the results shown in Fig.~\ref{fig4}. 
Here, for a fixed $g(v_{BL})$ value, the inflationary predictions are calculated 
   for various values of $\xi$ taken in Fig.~\ref{fig4}. 
We have used $m_{Z^\prime}=3$ TeV. 
The diagonal solid lines correspond to
  $g(v_{BL})= 0.0184$, $0.0216$, $0.023$, $0.026$, and $0.0425$.
}
\label{fig8}
\end{figure}

Finally, we show in Fig.~\ref{fig8} the contour plots (diagonal solid lines) 
   for the inflationary predictions with various fixed $g(v_{BL})$ values, 
   along with the results shown in Fig.~\ref{fig4}.   
Here, for a fixed $g(v_{BL})$ value, we calculate the inflationary predictions 
   from the effective inflaton potential for various values of $\xi$. 
We have used $m_{Z^\prime}=3$ TeV
The diagonal solid lines correspond to 
   $g(v_{BL})= 0.0184$, $0.0216$, $0.023$, $0.026$, and $0.0425$ from left to right.

\section{Reheating after inflation}
Any successful inflation scenario requires the transition to the Standard Big Bang Cosmology after inflation.
This happens via the decay of inflaton into the SM particles during the era inflaton is oscillating  
   around it potential minimum, and the decay products then reheat the universe. 
We estimate the reheating temperature after inflation $T_R$ 
   by  using $\Gamma  = H = T_R^2 \sqrt{\frac{\pi^2}{90} g_*}$,  
   where $\Gamma$ is the inflaton decay width, and $g_*$ is the effective degrees of freedom 
   for relativistic SM particles when the reheating occurs, so that   
\begin{eqnarray}
    T_R \simeq 0.2 \ (100/g_*)^{1/4} \sqrt{\Gamma M_P} .  
\label{TR}
\end{eqnarray} 
 From the success of big bang nucleosynthesis, we take a model-independent  lower bound on 
   the reheating temperature as $T_R \gtrsim 1$ MeV.

As we discussed in the previous section, the inflaton is much lighter than the $Z^\prime$ boson 
  and right-handed neutrinos. 
Thus, its decay width to the SM particles through off-shell process mediated by the heavy particles 
   is too small to satisfy the lower bound $T_R \gtrsim 1$ MeV.  
An efficient reheating process is possible when the inflaton in general has 
   a coupling with the SM Higgs doublet ($H$) such as
\bea
{\cal L} \supset  - \lambda^\prime \left(H^{\dagger} H  \right)  \left( \varphi^{\dagger} \varphi  \right)  . 
\label{H-phi}
\eea 
Although this coping is crucial for the reheating process, we assume $\lambda^\prime \ll 1$ 
    not to change our analysis for the $B-L$ model in the previous sections.   
Since the inflaton quartic coupling is very small and hence the inflaton is light, 
   it is most likely that the inflaton can decay to the SM particles 
   only through its mixing with the SM Higgs boson through the $\lambda^\prime$ coupling.  
After the $B-L$ and electroweak symmetry breakings, we diagonalize the scalar mass mass matrix of the form,  
\begin{eqnarray}
\begin{bmatrix} h \\ \phi \end{bmatrix}   =
\begin{bmatrix} \cos\theta &   \sin\theta \\ -\sin\theta & \cos\theta  \end{bmatrix} \begin{bmatrix} \phi_1 \\ \phi_2 
\end{bmatrix}  ,
\end{eqnarray} 
where $h$ is the SM Higgs boson, and  $\phi_1$  and $\phi_2$ are the mass eigenstates. 
The relations among the mass parameters and the mixing angle are the following: 
\bea
&2 v_{BL} v_{SM}  \lambda^\prime= ( m_h^2 -m_\phi^2) \tan2\theta,   \nonumber  \\
 &m_{\phi_1}^2 = m_h^2     - \left(m_\phi^2  - m_h^2 \right) \frac{\sin^2\theta}{1-2 \sin^2\theta} , \nonumber \\
 &m_{\phi_2}^2 = m_\phi^2 + \left(m_\phi^2 - m_h^2 \right) \frac{\sin^2\theta}{1-2 \sin^2\theta} \ .
\label{mixings} 
\eea
where $v_{SM}=246$ GeV is the Higgs doublet VEV, and $m_h$ and $m_\phi$ 
   are the masses for $h$ and $\phi$, respectively.

\begin{figure}[h]
\begin{center}
\includegraphics[scale=0.86]{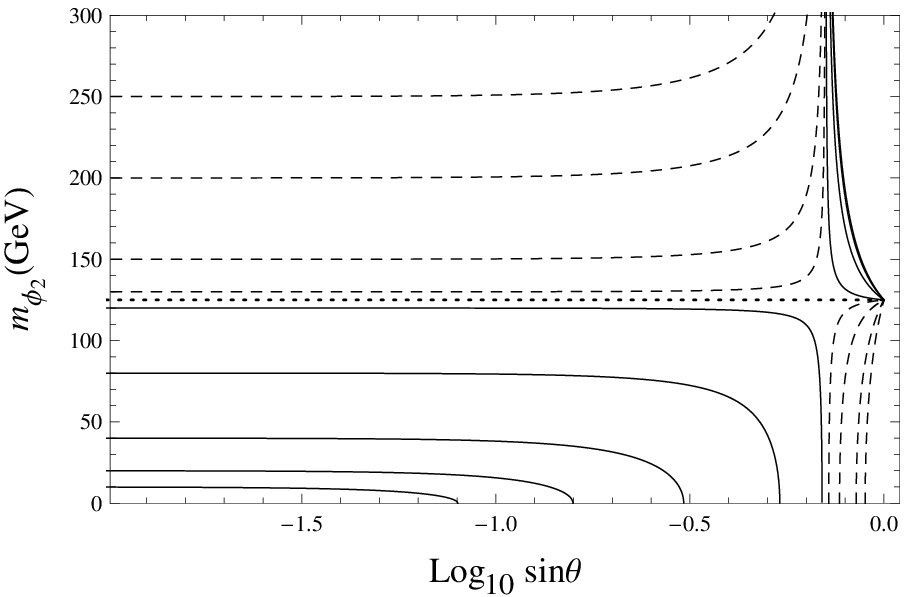} \; \; 
\includegraphics[scale=0.86]{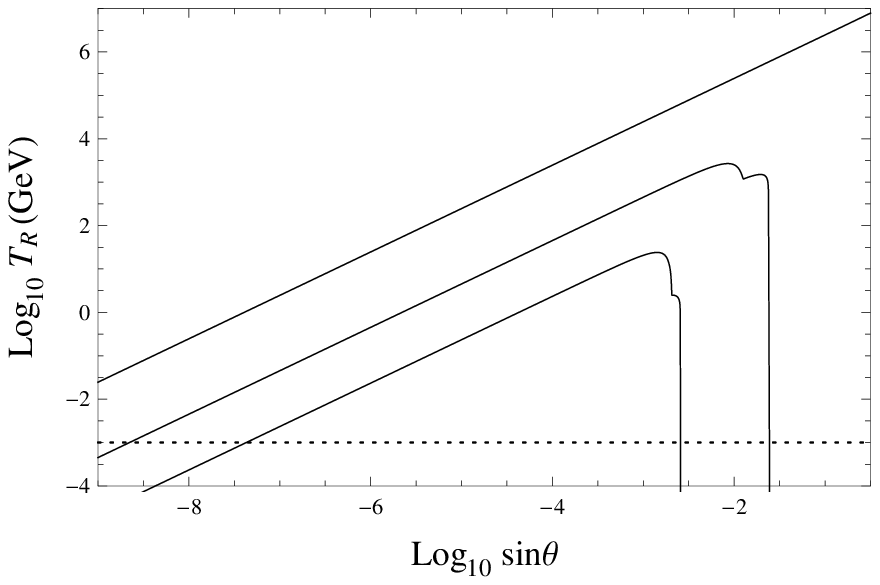} 
\end{center}
\caption{
Left: the mass eigenvalue $m_{\phi_2}$ as a function of $\sin\theta$ 
   for various $m_\phi$ values with $m_h=125$ GeV. 
We have used Eq.~(\ref{mixings}). 
Right: the reheating temperature as a function of $\sin\theta$ 
   for various values of $m_\phi=0.32$, $3.0$ and $130$ GeV (solid lines from bottom to top), 
   along with the dotted horizontal line for the lower bound of $T_R=1$ MeV. 
}
\label{fig9} 
\end{figure}

The left panel in Fig.~\ref{fig9} shows the mass eigenvalue $m_{\phi_2}$ as a function of $\sin\theta$ 
   for  various $m_\phi$ values with $m_h=125$ GeV.  
In the $\theta=0$ limit, $m_{\phi_2}=m_\phi$, while $m_{\phi}=m_h$ for $\theta=\pi/2$.  
Although in this plot we show the results by using  Eq.~(\ref{mixings}),  
  we only consider the case with $\theta \ll1$ as mentioned above, 
  otherwise our results obtained in the previous sections are changed 
  in the presence of a sizable $\lambda^\prime$. 
For $\theta \ll1$, the mass eigenstate $\phi_2$ ($\phi_1$) is almost identical to $\phi$ ($h$).

The inflaton can decay to the SM particles through the mixing with the SM Higgs boson. 
We calculate the inflaton decay width as 
\bea 
   \Gamma_{\phi_2} = \sin^2\theta \times  \Gamma_h(m_{\phi_2}) ,
\eea
where $\Gamma_h(m_{\phi_2})$ is the SM Higgs boson decay width 
   if the SM Higgs boson mass were $m_{\phi_2}$. 
The reheating temperature is then evaluated by Eq.~(\ref{TR}).  
For various inputs of $m_\phi$, we show the resultant reheating temperature 
  in the right panel of Fig.~\ref{fig8}. 
The solid lines from bottom to top denote the results for $m_\phi=0.32$, $3.0$ and $130$ GeV, respectively, 
  along with the dotted horizontal line for the lower bound of $T_R=1$ MeV. 
The sharp drop for each solid line corresponds to the fact that $m_{\phi_2}$ becomes zero 
  for a certain $\sin \theta$ value as shown in the left panel. 
We find that the universe is sufficiently heated up for a mixing angle in the suitable range shown in Fig.~\ref{fig8}.

Finally, we check the theoretical consistency of our analysis. 
When we introduce the coupling in Eq.~(\ref{H-phi}), the beta function of the inflaton quartic coupling 
   is modified to 
\bea 
   16 \pi^2  \beta_\lambda \to 16 \pi^2 \beta_\lambda + 2 \lambda^{\prime 2}.  
\eea
In order not to change our results in the previous sections by the introduction of $\lambda^\prime$,  
    $\lambda^{\prime 2}$  should be negligibly small in the beta function. 
We then impose a condition 
\bea 
   \frac{\lambda^{\prime 2}}{48 g^4}   \ll 1.  
 \label{cond}
\eea 
For $m_\phi^2 \ll m_h^2$, we obtain from Eq.~(\ref{mixings}) 
\bea 
   \lambda^\prime \simeq \frac{m_h^2}{v_{BL} v_{SM}}   \theta 
     \simeq  \frac{2 \; g \; m_h^2}{m_{Z^\prime} v_{SM}}   \theta  
     \simeq 4 \times 10^{-2} \; g \; \theta, 
\eea 
where we have used $m_{Z^\prime}=3$ TeV, $m_h=125$ GeV and $v_{SM}=246$ GeV. 
Combining this equation with Eq.~(\ref{cond}),  we obtain 
\bea 
    g \gg 6 \times 10^{-3} \; \theta .
\eea
 From Figs.~\ref{fig5}, \ref{fig7} and \ref{fig9}, we can see that this condition is satisfied 
   for a large potion of the parameters space.

\section{Conclusions}
Inflationary universe is the standard paradigm in modern cosmology, 
   which not only solves the problems in the Standard Big Bang Cosmology, 
   but also provide the primordial density fluctuations necessary 
   for generating the large scale structure of the present universe. 
As a simple and successful inflationary scenario, we have considered 
   the $\lambda \phi^4$ inflation with non-minimal gravitational coupling. 
With a suitable strength of the non-minimal coupling, the inflationary predictions 
  of this scenario becomes perfectly consistent with the Planck 2015 results.

It is more interesting if  the inflaton can also play some crucial role in particle physics. 
We have considered the general Higgs model with the gauge and Yukawa interactions 
   with the spontaneous gauge symmetry breaking.  
In the presence of the non-minimal gravitational coupling, the Higgs field can also play the role of inflaton.  
The analysis with the Higgs potential at the tree-level leads to the inflationary predictions 
  consistent with the cosmological observations. 
However, once we take quantum corrections, the effective inflaton potential most likely becomes unstable. 
This is because the inflaton quartic coupling is extremely small in a large portion of the parameters space  
    and the effective potential is controlled by the gauge and Yukawa couplings 
    independently of the quartic coupling. 
In the renormalization group improved effective potential, we see that the running quartic coupling 
    becomes negative in the vicinity of the initial inflaton value, indicating the instability of the effective potential. 
In order to avoid this problem, we have imposed the stability condition of vanishing the beta function 
    of the inflation quartic coupling.  
This condition leads to a non-trivial relation between the gauge and fermion masses. 
Since the renormalization group evolution of the inflaton quartic coupling is mainly controlled by 
   the gauge and Yukawa coupling, the inflation mass at low energy is determined by the couplings. 
Therefore, the mass spectrum of the gauge boson, fermion and inflation shows a non-trivial relation.

Since the inflaton potential is modified from the tree-level one,  
    the inflationary predictions are altered from those obtained by the tree-level analysis. 
Although the prediction of the tensor-to-scalar ratio remains the same 
    under the condition of the vanishing beta function,  
   the predictions for the scalar spectral index and the running of the spectral index can be significantly altered. 
The fact that the effective potential is controlled by the gauge and Yukawa couplings 
   implies a correlation between the inflationary predictions and the particle mass spectrum. 
Therefore, the observables at the gauge symmetry breaking scale correlate 
   with the inflationary predictions which determined by physics at an extremely high energy 
   compared to the gauge symmetry breaking scale.

By taking the minimal $B-L$ extension of the Standard Model as a simple example, 
   we have shown such a non-trivial relation in the particle mass spectrum 
   driven by the stability condition of the effective inflaton potential.  
We also have calculated the inflationary predictions from the effective potential 
   and found their dependence of the $B-L$ gauge coupling. 
Therefore, the new particle mass spectrum of the $B-L$ model, once observed, 
   has an implication to the inflationary predictions. 
On the other hand, more precise measurements of the inflationary predictions  
   yield a constraint on the $B-L$ particle mass spectrum.

For completeness, we have also investigated reheating after inflation. 
Since the inflation is lighter than the $Z^\prime$ boson and the right-handed neutrinos, 
  its reheating process through the heavy particles are not efficient, and the resultant 
  reheating temperature is too low to be consistent with the bound from big bang nucleosynthesis. 
We then introduce a coupling between the inflaton and the Standard Model Higgs doublet. 
Through the mixing with the Standard Model Higgs boson, the inflaton can decay into 
   the Standard Model particles and the universe can be heated up with a sufficiently high reheating temperature. 
We have found that  this happens with a sufficiently small coupling between the inflaton and the Higgs doublet  
   and such a small has essentially no effect on our analysis 
   for the particle mass spectrum and the inflationary predictions.

\section*{Acknowledgements}
This work is supported in part by the United States Department of Energy Grant, No. DE-SC 0013680.


\end{document}